# Probing the X-ray Spectra of Alpha Coronae Borealis

B. A. Korany[1,2]

Email: badiekorany@yahoo.com, baewiss@uqu.edu.sa

[1]Department of Physics, Faculty of Applied Science, Umm Al-Qura University, Saudi Arabia

[2]Department of Astronomy, National Research Institute of Astronomy and Geophysics (NRIAG), 11421 Helwan, Cairo, Egypt

**Abstract:** We present a detailed spectral and timing analysis of X-ray observations of the brightest eclipsing systems Alpha Coronae Borealis taken by XMM-Newton. We got from the thermal plasma model metal the abundances of some elements (O, Mg, Si, and Fe) and an emission line at 1.3 Kev from the simple Gaussian line profile. From the light curves, there is a strong active region at the lower left in the maps, near the limb of the G component, and increasing in some parts which means a band diagonally running across the G star disk from lower left to upper right, close to the projected center.

**Keywords**: Eclipsing Systems: X-ray: Alpha Coronae Borealis: α CrB

1- **Introduction**

Alpha Coronae Borealis (α CrB) is one of the brightest eclipsing systems, so a lot of studies were interested in this system in different energy bands. It was discussed early by Stebbins in 1914 in optical band. Tomkin (1986) gives a detailed optical information for this system as; it is a 17.4 day eclipsing binary, the spectral types of its components is A and G with different mass, the G-type component is dwarf star in 17.4 day orbital period, and the system orbital property is i=88.2, $r_p$ =0.07±0.007, $r_s$ = 0.021±0.001. Schmitt in 2016 used the TIGRE 1.2 m telescope observation to get a new radial velocity curve, apsidal motion, and the alignment of rotation and orbit axes. The apsidal motion rate $0.035 < \dot{\omega} < 0.054$ degrees/year and the apsidal motion period (Paps) is in the range 6600 yr < Paps < 10 600 yr, while the apsidal motion values results in a difference 7.2 s < ∆ Psp = Ps − Pp < 11.6 s (Schmitt et.al 2016).

First X-ray observation of this system was by ROSAT in 1993 (Schmitt and Kurster, 1993). Schmitt and Kurster (1993 & 1994) reconstructed a rough surface X-ray map, applying a maximum likelihood method. The resulting map revealed a patchy surface coverage of X-ray emitting material. The X-ray light curve of α CrB shows a total X-ray eclipse during the secondary optical minimum, with the G star behind the A star. The totality of the eclipse demonstrates that the A-type component in α CrB is X-ray dark and that the x-ray flux arises exclusively from the later-type companion (Schmitt and Kurster 1993). Güdel et al. (2003) used X-ray data observed by XMM-Newton during a total X-ray eclipse of this binary system and studied its tomography of a stellar X-ray corona. They calculated the binary orbital elements (a, e, i, ω: $2.981 \times 10^{12}$ cma, 0.370, 88.2º, 311.0º respectively) and the stellar radii are $R_A$=3.04 $R_\odot$ and $R_G$ =0.09 $R_\odot$.

We report in this paper on spectral and timing analysis of X-ray observations of the eclipsing system α CrB , taken by XMM-Newton observations. We organized the paper as follows: The X-Ray observations and data reduction are presented in section 2, section 3 is devoted to the spectral analysis, while the results are summarised and concluded in section 4.

2- **X-Ray Observation and data reduction**

We used XMM-Newton observations for α CrB from the XMM-Newton archive, the observing log is given in Table 1. All EPIC cameras were operated in the small-window mode with the thick filter inserted in order to suppress the strong optical flux from the primary A star (Güdel et al 2003). We made use of the data from the two EPIC MOS (Turner et al. 2001) cameras and the EPIC PN camera (Strüder et al. 2001). The raw data were processed with the EPIC pipeline chains of the Science Analysis System (SAS) software version 9.0. Some bad time intervals characterized by high background events (so-called soft-proton flares) were rejected by creating light-curves for the observations (PN, MOS1, and MOS2), which are best visible above 10 keV. To optimize the signal-to-noise ratio in the light curve and to suppress large background contributions and contributions from warm pixels in the very soft range, because of the poverty of the hard x-ray photons of this object, the only soft X-Ray band extracted exclusively in the energy range 0.30-2.0 keV for the PN and 0.14-2.5 keV for the MOS1&MOS2 cameras. The spectrum of the star was extracted using circular extraction regions centered on the star, with radii of approximately 25" for both the MOS and PN cameras. In order to utilize the $\chi^2$ technique, the X-ray spectra were rebinned to contain at least 20 counts in each spectral bin using *grppha* command. All the spectra were subsequently analyzed using the **Xspec** software (v12).

For the timing analysis, we extracted the source and background light curve by using the *evselect* task of SAS software version 9.0 . The light curves obtained corrected to account for a number of effects which an impact in the detection efficiency can have, like vignetting, bad pixels, PSF variation and quantum efficiency, as well as to account for time-dependent corrections within an exposure, like dead time and GTIs, by using the SAS task epiclccorr. Because in the short time periods x-ray binaries the arrival time of a photon is shifted as is it would have been detected at the barycenter of the solar system (the center of mass) instead at the position of the satellite. In this way, the data are comparable. We used the SAS barycen to correct these timestamps to the Earth's barycenter for each event. The Xronos program package used to make a timing analysis (producing a binned light curve, calculating a power spectrum, searching for periodicities etc.).

### Table 1  Observing log for α CrB with XMM-Newton

| Instrument | Mode | Filter | Date @Exp. Start Time | Date @ Exp. End Time |
|---|---|---|---|---|
| MOS1 | Small-window | THICK FILTER | 2001-08-27 @04:42:44 | 2001-08-27 @15:42:18 |
| MOS2 | Small-window | THICK FILTER | 2001-08-27 @04:42:44 | 2001-08-27 @15:42:18 |
| PN | Small-window | THICK FILTER | 2001-08-27 @04:58:46 | 2001-08-27 @15:43:09 |

3- **X-Ray Data Analyses**

The first model used to fit the spectra is a single temperature variable- abundance thermal plasma model, in this model an emission spectrum from hot diffuse gas based on the model calculations of Mewe and Kaastra with Fe L calculations. The model includes line emissions from several elements. The He element in the model fixed at cosmic, and the normalization calculated from following equation:

$$\frac{10^{-14}}{4\pi[D_A(1+z)]^2}\int n_e n_H dv \quad (1)$$

Where $D_A$ is the angular diameter distance to the source in cm, $n_e$ & $n_H$ are the densities in $cm^{-3}$ of the electron and hydrogen respectively ( Liedahl et al 1995) (in Xspec software this model called vmekal). Figures 1 and 2 explain the fit for PN and MOS, while Figure 3 is the fitting of the three detectors (PN, MOS1, and MOS2 respectively). The elements of metal abundance included in the model are C, N, O, Ne, Na, Mg, Al, Si, S, Ar, Ca, Fe and Ni. We set

O, Ne, Mg, Si, S, and Fe are free parameters in the fit. We got a best fit for this single model, $\chi^2$/odf is 141.16/121. The output temperature is 0.46±0.02, while the output values of the abundance of the elements O, Mg, S, and Fe are 0.25±0.05, 0.44±0.12, 0.52±0.27 and 0.26±0.02 respectively. The X-ray luminosity evaluated from 0.3-10.0 Kev is $5.3 \times 10^{28}$ erg s$^{-1}$, which agrees with the luminosity reported Schmitt & Kurster 1993 and Gudel, et al 2003. By adding the A photo-electric absorption due to the hydrogen column density no significant change in the fitting (the results are summarized in Table 2). The simple Gaussian line profile, which calculated from the equation

$$A(E) = K \frac{1}{\sigma\sqrt{(2 \times \pi^2)}} e^{\left(\frac{-(E-E_t)^2}{2\sigma^2}\right)} \quad (2)$$

(here $E_t$ is line energy in keV, $\sigma$ is the line width in keV and K is the total photons/cm$^{-2}$/s in the line), was applied the fitting was Improved ( $\chi^2$ /odf = 123.75/ 107) (the fitting of the thermal plasma model with simple gaussian line for PN data shown in figure 5) and we got an emission line at 1.3 Kev (all the parameter in table 2).

The produced binned light curve for the three detectors (PN, MOS1, and MOS2) explains that the data cover the complete second eclipse (Fig 6-9). The gross shape of the ingress and egress portions of the curve is symmetric, the total ingress and egress durations are 2.74 and 1.66 hrs. respectively. During ingress and egress a sequence of fast drops/rises in flux and several flux plateaus are evident. We can note from the light-carve figures a strong asymmetry in all light curves, and the individual steep flux decreases and increases. We can see from figures 5 to 7 a strong active region at the lower left in the maps, near the limb of the G component, the lower right part which in increasing means a band diagonally running across the G star disk from lower left to upper right, close to the projected center. The upper right part has the same features of the lower part, but weaker band displaced somewhat to the right, and to the bright source of the upper right secondary limb. The banded structure from lower left to upper right may partially be an artifact since the ingress light curve decays relatively smoothly below feature the lower right part.

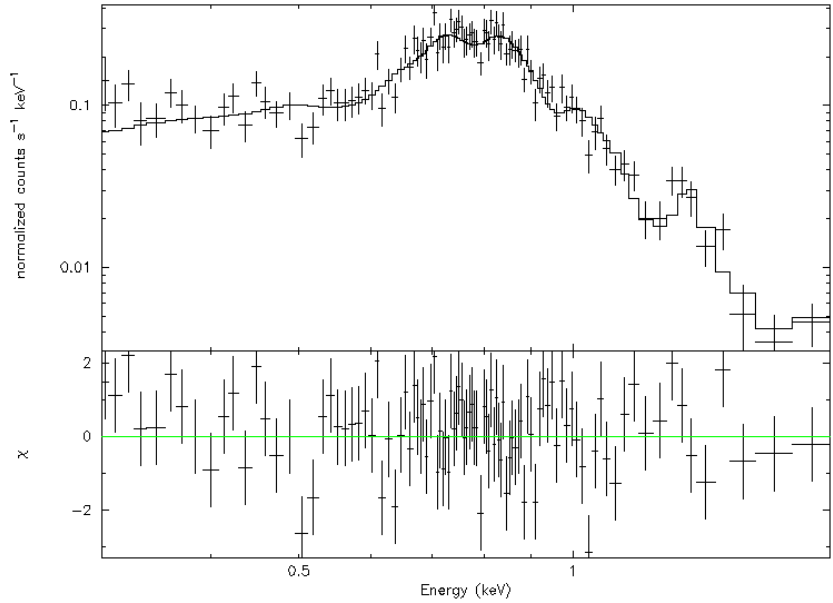

Fig1: The EPIC PN spectrum which fitted using single thermal plasma model, with $\chi^2$ test.

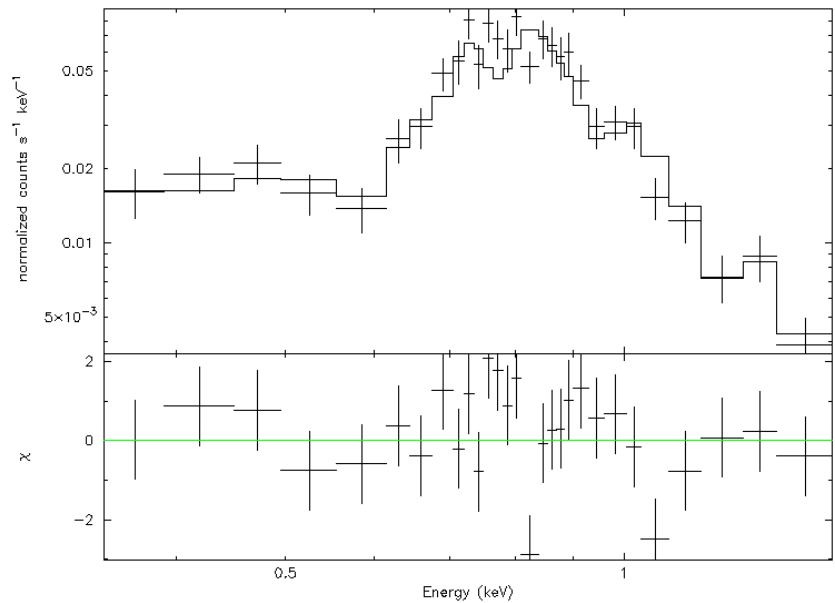

Fig2: The EPIC MOS1 spectrum which fitted using single thermal plasma model, with $\chi^2$ test.

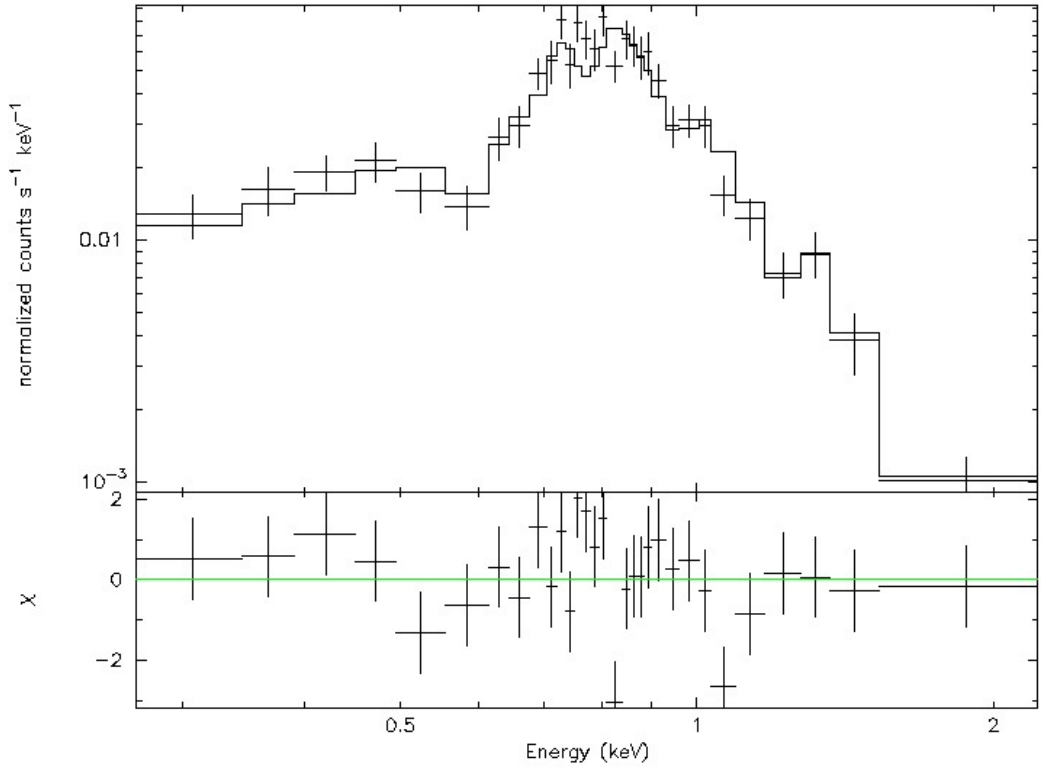

Fig3: The EPIC MOS2 spectrum which fitted using single thermal plasma model, with $\chi^2$ test.

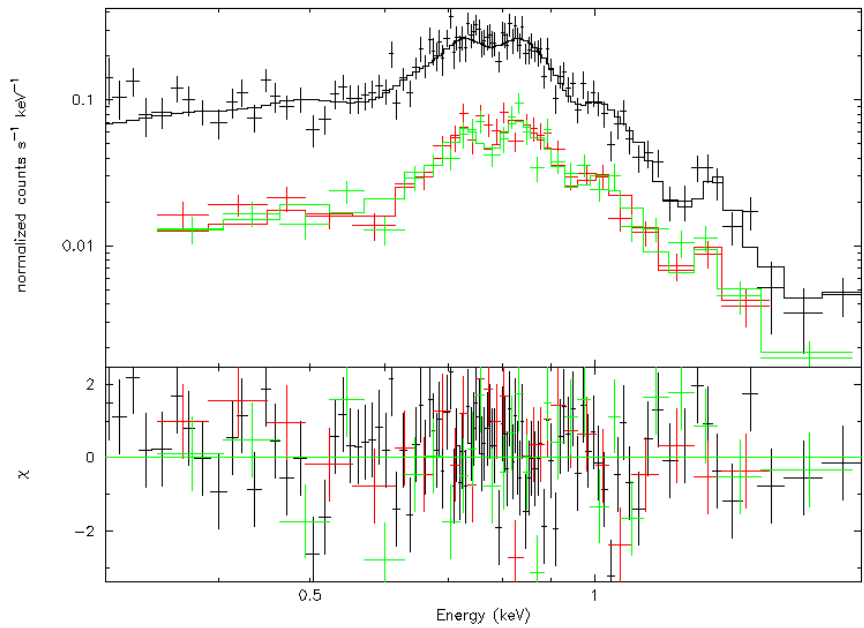

Fig4: The EPIC PN, MOS1 and MOS2 spectra which fitted using single thermal plasma model, with $\chi^2$ test.

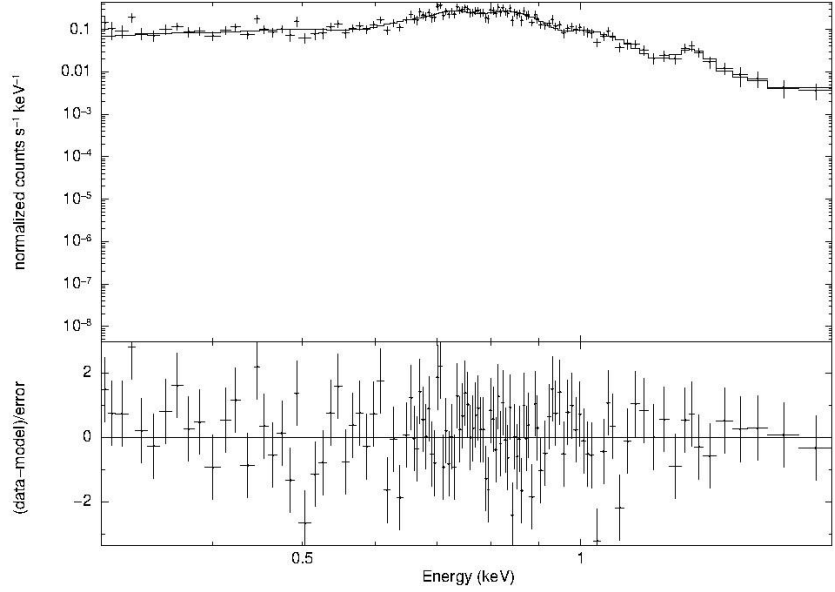

Fig5: The EPIC PN spectrum which fitted using single thermal plasma model and single gaussian line, with $\chi^2$ Test.

Table 2. Summarizing the parameters of the output fitting

| Parameter | Thermal plasma model | Thermal plasma model+ A photo-electric absorption | Thermal plasma model+ Gaussian line |
|---|---|---|---|
| Kt | 0.46 ±0.02 Kev | 0.46 ±0.02 Kev | 0.47 ±0.02 Kev |
| O | 0.25 ±0.05 | 0.27±0.06 | 0.33±0.06 |
| Mg | 0.44±0.12 | 0.47±0.13 | 0.54±0.16 |
| Si | 0.52±0.17 | 0.56±0.28 | 0.37±0.19 |
| Fe | 0.26±0.02 | 0.28±0.03 | 0.31±0.06 |
| $\chi^2$/odf | 141.16/121 | 140.65/111 | 123.75/107 |
| Normalization | $2.5 \times 10^{-4}$ | $2.4 \times 10^{-4}$ | $2.6 \times 10^{-5}$ |

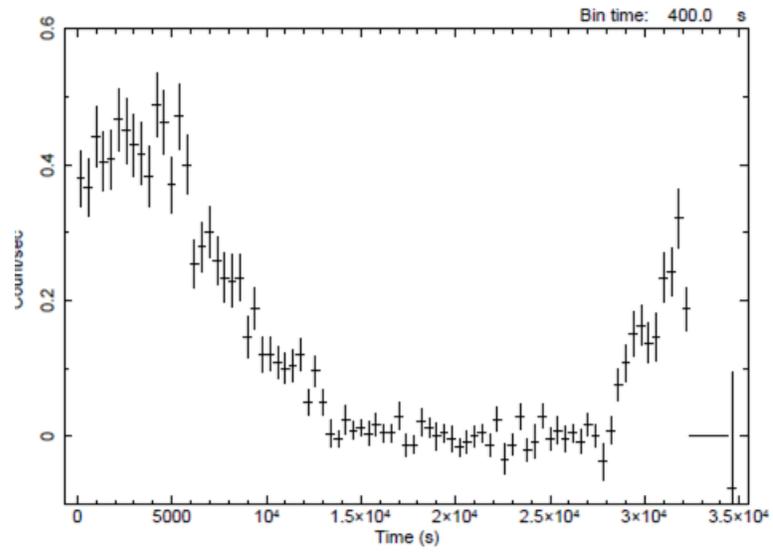

Fig6: The X-ray light curve for PN data

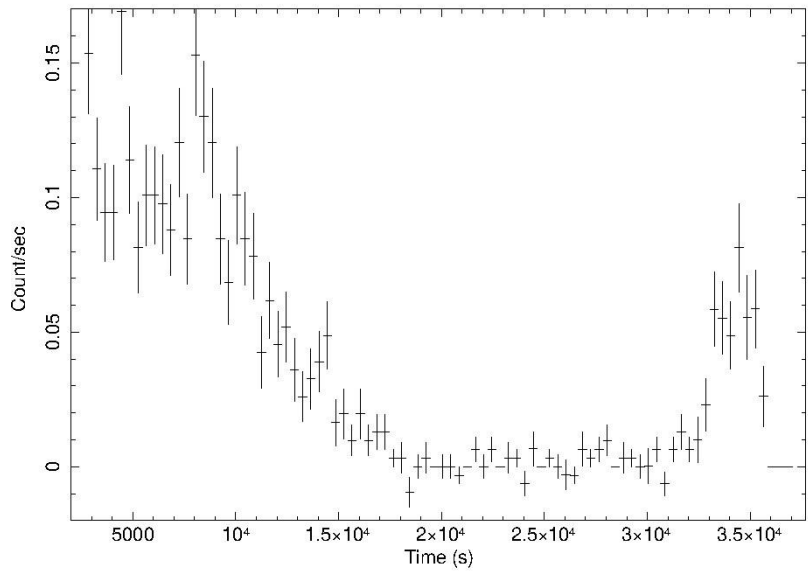

Fig7: The X-ray light curve for MOS1 data

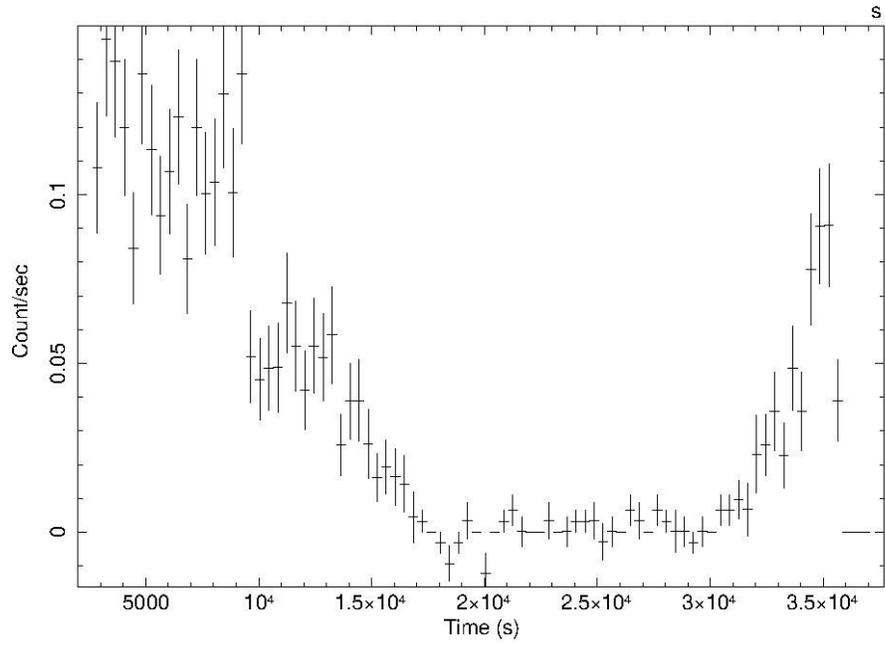

Fig8: The X-ray light curve for MOS2 data

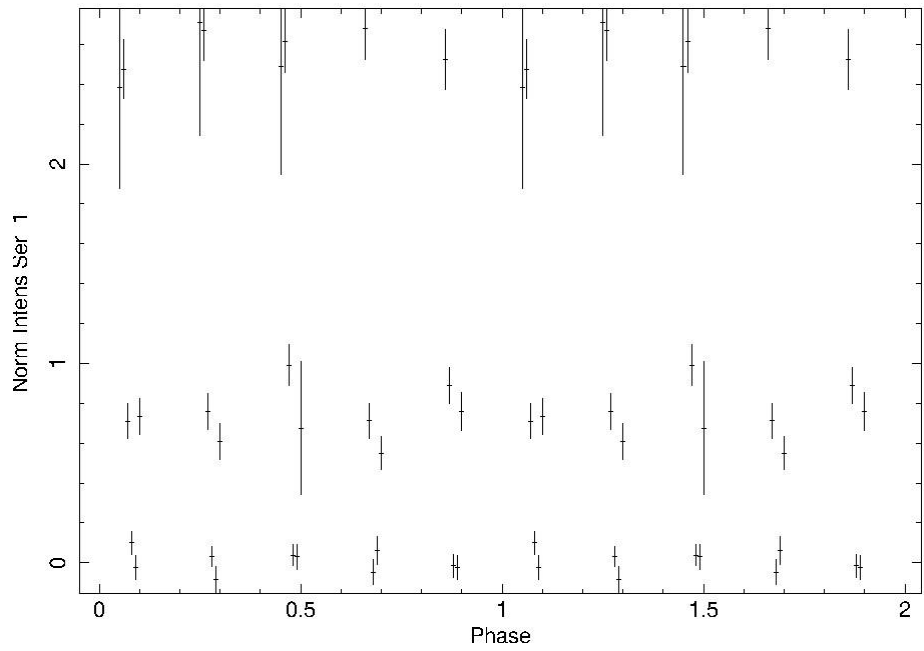

Fig9: Phase of the eclipsing binary

## 4- Discussion and conclusions

A detailed spectral and timing analysis (EPIC) has been performed for XMM-Newton observation to the alpha Coronae Borealis (α CrB) system. The soft X-Ray range (0.3-2.0 Kev) is described by three models. From these models we got the best-fit temperature in the range 0.45 to 0.47 kev, while the abundances of the elements O, Mg, Si and Fe which have the range 0.25 to 0.33, 0.44 to 0.54, 0.37 to 0.56 and 0.26 to 0.31 respectively, are low compared to solar photospheric values. By using Guessing model there is an emission line at 1.3 Kev. From the timing analysis there is a strong active region at the lower left in the maps, near the limb of the G component, and increasing in some parts means a band diagonally running across the G star disk from lower left to upper right, close to the projected center. Some parts indicated that a weaker band displaced somewhat to the right, and to the bright source of the upper right secondary limb. The banded structure from lower left to upper right may partially be an artifact since the ingress light curve decays relatively smoothly below feature of the lower right part.

## References

1. Liedahl, D.A., Osterheld, A.L., and Goldstein, W.H. 1995, ApJL, 438, 115

2. M. Güdel, K. Arzner, M. Audard and R. Mewe 2003, A&A, 403, 155-171

3. Schmitt, J. H. M. M., & Kürster, M. 1993, Science, 262, 215

4. Schmitt1, J. H. M. M., Schröder, K.-P., Rauw, G., Hempelmann, Mittag, A, M., González-Pérez1 J. N., Czesla1, S., Wolter, U., and Jack, D. A&A, 2016, 586, A104

5. Strüder, L.; Briel, U.; Dennerl, K.; …..et al , 2001, A&A, 365, L18

6. Turner, M. J. L., Abbey, A., Arnaud, M., et al., 2001, A&A, 365, 27.

7. Tomkin, J. and Popper, D. M, 1968 AJ 92, 6.